\documentclass[12pt]{article}
\begin{document}
\begin{center}
{\bf RELATIVISTIC WAVE EQUATION FOR FIELDS WITH TWO MASS AND SPIN
STATES}\\
\vspace{5mm}
 S.I. Kruglov \footnote{E-mail: skrouglo@utm.utoronto.ca}\\
\vspace{5mm} \textit{University of Toronto at Scarborough,
Physical and Environmental Sciences Department, 1265 Military
Trail, Toronto, Ontario, Canada M1C 1A4}
\end{center}

\begin{abstract}
I suggest wave equations for the scalar, pseudoscalar, vector, and
pseudovector fields with different masses for spin zero and one
states. Tensor, matrix, and quaternion formulations of fields with
two mass and spin states are considered. This is the
generalization of the Dirac-K\"ahler equation on the case of
different masses of fields with spin one and zero. The equation
matrices obtained are simple linear combinations of matrix
elements in the 16-dimensional space. Spin projection operators
and solutions of equations (for spin one) in the form of
matrix-dyads are obtained. The canonical quantization of fields
under consideration is studied. The anomalous interaction of the
scalar, pseudoscalar, vector, and pseudovector fields with the
external electromagnetic field is considered. Three constants
which characterize the anomalous magnetic moment and quadrupole
electric moment of a particle are introduced.
\end{abstract}

\section{Introduction}

The modern high energy physics require the consideration of fields
with higher spins. So, supersymmetrical (SUSY) field models
introduce superpartners of ordinary particles which are high spin
particles. Cosmological models also deal with fields possessing
different spins.

The theory of relativistic wave equations describing particles
with arbitrary spins has a long history \cite{Fierz}, but is not
complete yet. The simple high spin fields are fields with spin
one. Ordinarily, spin-1 particles are described by Proca
\cite{Proca} or Petiau-Duffin-Kemmer equations \cite{Petiau},
\cite{Duffin}, \cite{Kemmer}. Anomalous interactions of vector
particles with electromagnetic fields were considered in
\cite{Taketani}, \cite{Corben}, \cite{Joung}. Other schemes to the
theory of spin one particles were developed by
\cite{Stueckelberg}, \cite{Lee}.

K\"ahler \cite{Kahler} considered an equation for inhomogeneous
differential forms which is equivalent to a system of scalar,
vector, antisymmetric tensor, pseudovector and pseudoscalar fields
\cite{Kruglov1}, \cite{monogr}. Now, the K\"ahler equation (called
Dirac-K\"ahler equation), in the framework of differential forms
is intensively used for description of quarks on the lattice
\cite{Becher}. Note that some authors \cite{Ivanenko} considered
equations for the system of antisymmetric tensor fields which are
equivalent to the K\"ahler equation before the appearance of
\cite{Kahler}. It should be mentioned that the Lagrangian of a
system of scalar, vector, antisymmetric tensor, pseudovector and
pseudoscalar fields (equivalent to Dirac-K\"ahler fields) is
invariant under the internal symmetry group $SO(4,2)$ (or locally
isomorphic group $SU(2,2)$) \cite{Kruglov1}, \cite{monogr}.

In this paper, we generalize the Dirac-K\"ahler equation on the
case that scalar, pseudoscalar fields having different masses,
$m_0$, compared with vector, pseudovector fields, $m$. At the
particular case $m_0=m$, one arrives at the Dirac-K\"ahler fields.

The paper is organized as follows: In Section 2, we introduce the
field equations which are the generalization of the Dirac-K\"ahler
equations. It is prove that fields introduced describe scalar,
pseudoscalar fields with mass $m_0$ and vector, pseudovector
fields with mass $m$. We show that introduced 16-component system
of equations can be represented as two 8-component independent
sub-systems with self-dual, and antiself-dual antisymmetrical
tensors. At parity transformations, these systems are converted to
each other. In Section 3, we construct the matrix form of
16-component relativistic wave equation. It is shown that matrices
of the equation obey the Dirac commutation relation. This equation
also contains two projection operators which are connected with
two possible mass states. Section 4 is devoted to the matrix form
of 8-component equations. In the case of equal field masses,
$m_0=m$, the 8-component matrices also satisfy the Dirac
commutation relation. In section 5, we study the Lorenz covariance
of equations under consideration. The one-parameter symmetry group
of the Lagrangian and the corresponding conserve current are
investigated here. The quaternion form of fields is constructed in
section 6. This quaternion form includes different particular
cases of field schemes such as Proca, Maxwell, etc. In section 7,
spin projection operators and solutions of equations for spin one
in the form of matrix-dyads are found. The quantization of fields
is carried out in Section 8. In section 9, we consider the
possible anomalous electromagnetic field interaction. Section 10
contains the discussion. The appendix is devoted to the
computation of products of 16-component matrices.

The system of units $\hbar =c=1$, $\alpha =e^2/4\pi =1/137$, $e>0$
is used.

\section{Field equations}

Now, I suggest new field equations for the set of bosonic fields
which have two spin one states and two spin zero states with
differen masses. In particular cases, these equations are
transformed to one considered in literature.

Let us introduce and investigate the system of equations for free
scalar $\varphi (x)$, pseudoscalar $\widetilde{\varphi }(x)$,
vector $\varphi _\mu (x)$, pseudovector $\widetilde{\varphi }_\mu
(x)$ and antisymmetric tensor $\varphi _{\mu \nu }(x)$ fields:
\begin{equation}
\partial _\nu \varphi _{\mu \nu }(x)-\partial _\mu \varphi (x)+m^2\varphi _\mu
(x)=0 ,\hspace{0.2in}\partial _\nu \widetilde{\varphi }_{\mu \nu
}(x)-\partial _\mu \widetilde{\varphi }(x)+m^2\widetilde{\varphi
}_\mu (x)=0  ,\label{1}
\end{equation}
\begin{equation}
a\partial _\mu \varphi _\mu (x)-\varphi (x)=0
,\hspace{0.3in}a\partial _\mu \widetilde{ \varphi }_\mu
(x)-\widetilde{\varphi }(x)=0 , \label{2}
\end{equation}
\begin{equation}
\varphi _{\mu \nu }(x)=\partial _\mu \varphi _\nu (x)-\partial
_\nu \varphi _\mu (x)-\varepsilon _{\mu \nu \alpha \beta }\partial
_\alpha \widetilde{\varphi } _\beta (x) , \label{3}
\end{equation}
where $\partial_\mu=(\partial_m ,-i\partial_0)$, $x_\mu=(x_m
,ix_0)$; $x_0$ is the time, and
\begin{equation}
\widetilde{\varphi }_{\mu \nu }(x)=\frac 12\varepsilon _{\mu \nu
\alpha \beta }\varphi _{\alpha \beta }(x) \label{4}
\end{equation}
is the dual tensor, $\varepsilon _{\mu \nu \alpha \beta }$ is an
antisymmetric tensor Levy-Civita; $\varepsilon _{1234}=-i$; $a$
and $m$ are parameters which are connected with the masses of
scalar, pseudoscalar and vector, pseudovector fields. Equations
(1)-(4) are the generalization of the tensor form of the
Dirac-K\"ahler equation \cite{Kahler}. At $a=1$, we come to
Dirac-K\"ahler equations \cite{Kruglov1} (see also \cite{monogr});
the case $\widetilde{\varphi }(x)=0$ and  $\widetilde{ \varphi
}_\mu (x)=0$ was considered in \cite{Kruglov2}; at $a=1$,
$\widetilde{\varphi }(x)=0$, $\widetilde{ \varphi }_\mu (x)=0$, we
arrive at the Stueckelberg equations \cite{Stueckelberg} (see also
\cite{Kruglov3}); the case $a=0$, corresponding vector and
pseudovector fields, was investigated in \cite{Kruglov4}; in the
case $\varphi (x)=0$, $\widetilde{\varphi }(x)=0$, $\widetilde{
\varphi }_\mu (x)=0$ one arrives at the Proca equations
\cite{Proca}; at $a=1$, $\widetilde{\varphi }(x)=0$, $\widetilde{
\varphi }_\mu (x)=0$, $m=0$, we obtain generalized Maxwell's
equations investigated in \cite{Kruglov5}. So, field equations
(1)-(4) include many important cases of equations for bosonic
fields with different number of freedom degrees.

To clear up the physical meaning of constants $a$, $m$, one can
get from Eqs. (1)-(4) the second order equations for vector and
pseudovector fields
\[
\partial _\nu^2 \varphi _{\mu }(x)-(1-a)\partial _\mu\partial _\nu
\varphi_\nu (x) -m^2\varphi _\mu(x)=0 ,
\]
\vspace{-8mm}
\begin{equation}  \label{5}
\end{equation}
\vspace{-8mm}
\[
\partial_\nu^2\widetilde{\varphi}_{\mu}(x)-(1-a)\partial _\mu
\partial_\nu\widetilde{\varphi}_\nu (x)-m^2\widetilde{\varphi}_\mu (x)=0 .
 \]
Because two equations in (5) are identical, masses of vector and
pseudovector fields are the same. To obtain the masses of vector
(and pseudovector) fields, we write the equation for vector field
(5) in momentum space (see \cite{Kruglov2}):
\begin{equation}
 \left[\left(p_\lambda^2+m^2\right)\delta_{\mu\nu}-(1-a)p_\mu p_\nu
\right]\varphi_\nu (p)=0 , \label{6}
\end{equation}
where $p^2=p_\lambda ^2={\bf p}^2+p_4^2={\bf p}^2-p_0^2$. The
matrix
\begin{equation}
M=\left( p^2+m^2 \right) I_4-(1-a) \left( p_{\cdot }p\right),
\label{7}
\end{equation}
where the $I_4$ is the unit 4$\times $4-matrix and the matrix-dyad
$\left(p_{\cdot }p\right)$ has the matrix elements $ \left(
p_{\cdot }p\right) _{\mu \nu }=p_\mu p_\nu $, obeys the equation
\begin{equation}
\left(M-m^2 -p^2\right) \left[ M-m^2 -ap^2 \right] =0 . \label{8}
\end{equation}
It follows from Eq. (8) that the matrix $M$ possesses the
eigenvalues $\lambda _1=p^2+m^2 $ , $\lambda _2=ap^2+m^2 $. So,
Eq. (6) has nontrivial solutions at $\det M=0$ or $\lambda
_1=\lambda _2=0$. This leads to the mass spectrum of the vector
field $\varphi_\nu (x)$:
\begin{equation}
p^2=-m^2,\hspace{1.0in}p^2=-\frac{m^2}{a }. \label{9}
\end{equation}
We can draw a conclusion from Eq. (9) that one state of the field
$\varphi_\nu (x)$ corresponds to the mass $m$, and the second
state corresponds to the squared mass
\begin{equation}
m_0^2=\frac{m^2 }{a} . \label{10}
\end{equation}
The requirement that the masses are real ($m^2>0$, $m_0^2>0$), we
get the condition $a >0$. To identify the the mass states with the
spin states, one finds from Eqs. (1)-(4) the second order equation
for the scalar and pseudoscalar fields:
\[
\partial _\nu ^2 \varphi (x)-\frac{m^2}{a}\varphi (x) =0 ,
\]
\vspace{-8mm}
\begin{equation}  \label{11}
\end{equation}
\vspace{-8mm}
\[
\partial _\nu ^2 \widetilde{\varphi}(x) -\frac{m^2}{a}\widetilde{\varphi} (x)=0.
\]
From Eqs. (10), (11) we conclude that scalar $\varphi (x)$ and
pseudoscalar $\widetilde{\varphi}(x)$ fields with spin $s=0$
possess the mass $m_0$. Therefore, the vector and pseudovector
states with spin $s=1$ have the mass $m$. In the case of $a =1$,
states with spin $s=1$ and $s=0$ possess the same mass $ m$. For
arbitrary variable $a>0$ the system of equations (1)-(4) describes
fields which may have the vector and pseudovector states with the
mass $m$ and scalar and pseudoscalar states with the mass $m_0$.
The Dirac-K\"ahler equation \cite{Kahler} is obtained by setting
$m=m_0$ ($a=1$).

Introducing the variables
\[
M(x)=\frac 1{\sqrt{2}}\left( \varphi (x)-i\widetilde{\varphi
}(x)\right) , \hspace{0.3in}N(x)=\frac 1{\sqrt{2}}\left(\varphi
(x)+i\widetilde{\varphi}(x)\right) ,
\]
\[
M_\mu (x)=\frac 1{\sqrt{2}}\left( \varphi_\mu
(x)-i\widetilde{\varphi}_\mu (x)\right) , \hspace{0.3in}M_{\mu \nu
}(x)=\frac 1{\sqrt{2}}\left( \varphi _{\mu \nu }(x)-i
\widetilde{\varphi}_{\mu \nu }(x)\right) ,
\]
\[
 N_\mu (x)=\frac
1{\sqrt{2}}\left( \varphi_\mu (x)+i\widetilde{\varphi}_\mu
(x)\right) , \hspace{0.3in}N_{\mu \nu }(x)=\frac 1{\sqrt{2}}\left(
\varphi _{\mu \nu }(x)+i \widetilde{\varphi }_{\mu \nu }(x)\right)
,
\]
and adding and subtracting Eqs. (1)-(4), we obtain equations
\[
\partial _\nu M_{\mu \nu }(x)-\partial _\mu M(x)+m^2M_\mu (x)=0,
\hspace{0.3in}a\partial _\mu M_\mu (x)=M(x) ,
\]
\vspace{-8mm}
\begin{equation}
\label{12}
\end{equation}
\vspace{-8mm}
\[
M_{\mu \nu }(x)=\partial _\mu M_\nu (x)-\partial _\nu M_\mu
(x)-i\varepsilon _{\mu \nu \alpha \beta }\partial _\alpha M_\beta
(x) ,
\]
\[
\partial _\nu N_{\mu \nu }(x)-\partial _\mu N(x)+m^2N_\mu (x)=0 ,
\hspace{0.3in}a\partial _\mu N_\mu (x)=N(x) ,
\]
\vspace{-8mm}
\begin{equation}
\label{13}
\end{equation}
\vspace{-8mm}
\[
N_{\mu \nu }(x)=\partial _\mu N_\nu (x)-\partial _\nu N_\mu
(x)+i\varepsilon _{\mu \nu \alpha \beta }\partial _\alpha N_\beta
(x) .
\]
The tensor $M_{\mu \nu }(x)$ is self-dual tensor, $M_{\mu \nu
}(x)=-i\widetilde{M}_{\mu \nu }(x)$, and the tensor $N_{\mu \nu
}(x)$ is the antiself-dual tensor, $N_{\mu \nu
}(x)=i\widetilde{N}_{\mu \nu }(x)$.

The self-dual tensor $M_{\mu \nu }$ possesses $3$ independent
components and realizes the $\left( 1,0\right) $-representation of
the Lorentz group. The antiself-dual tensor $N_{\mu \nu }$ also
has $3$ independent components and is transformed under the
$\left( 0,1\right) $-representation of the Lorentz group. Eqs.
(12) and (13)) are $8$-component equations which describes eight
independent variables ($M(x)$, $M_\nu (x)$ , $M_{ab}(x)$),
($N(x)$, $N_\nu (x)$, $N_{ab}(x)$) are not invariant under the
parity transformations separately. At P-transformations, Eqs. (12)
are transferred into Eqs. (13) and vice versa. It is possible to
have Lagrangian formulation only for the whole system of equations
(12), (13) or (1)-(4). P-noninvariant equations (12) (or (13))
have no the Lagrangian formulation. Eqs. (1)-(4) realize the
$(0,0)\oplus (1/2,1/2)\oplus \left( 1,0\right) \oplus \left(
0,1\right) \oplus (1/2,1/2)\oplus (0,0)$-representation of the
Lorentz group, and are equivalent to equations (12), (13)).

\section{Matrix form of 16-component equations}

Now we obtain the matrix form of equations (1)-(4). To get the
relativistic wave equation in the matrix form, we introduce new
variables
\[
\psi _0 (x)=-\varphi (x) ,~~ \psi _\mu (x)=m\varphi _\mu (x) ,
~~\psi _{[\mu \nu ]}(x)=\varphi _{\mu \nu } (x) ,
\]
\[
 \widetilde{\psi }_\mu (x)=im\widetilde{ \varphi }_\mu (x) ,~~
\widetilde{\psi }_0 (x)=-i\widetilde{\varphi } (x) ,~~ e_{\mu \nu
\alpha \beta }=i\varepsilon _{\mu \nu \alpha \beta }~~
(e_{1234}=1).
\]
Then Eqs. (1)-(4) with the help of Eq. (10) are rewritten as
\[
\partial _\mu \widetilde{\psi }_\mu (x)+\frac{m_0^2}{m}\widetilde{\psi }_0 (x)=0 ,
\]
\[
\partial _\nu \psi _{[\mu \nu ]}(x)+\partial _\mu \psi _0 (x)+m\psi
_\mu (x)=0 ,
\]
\vspace{-8mm}
\begin{equation}  \label{14}
\end{equation}
\vspace{-8mm}
\[
\partial _\nu \psi _\mu (x)-\partial _\mu \psi _\nu (x)-e_{\mu \nu \alpha \beta
}\partial _\alpha \widetilde{\psi }_\beta (x)+m\psi _{[\mu \nu
]}(x)=0 ,
\]
\[
\partial _\mu \psi _\mu (x)+\frac{m_0^2}{m}\psi _0 (x)=0 .
\]
With $m_0=m$, we arrive at the Dirac-K\"ahler equations. It is
convenient to introduce the 16-component wave function
\begin{equation}
\Psi (x)=\left\{ \psi _A (x)\right\} ,\hspace{0.5in}A=0,\mu ,[\mu
\nu ], \widetilde{\mu },\widetilde{0} , \label{15}
\end{equation}
where $\psi _{\widetilde{\mu}}\equiv\widetilde{\psi}_\mu$, $\psi
_{\widetilde{0}}\equiv\widetilde{\psi}_0$. Let us introduce the
the elements of the entire algebra $\varepsilon ^{A,B}$
\cite{Bogush} with dimension $16\times 16$. The product of these
matrices and their matrix elements are given by
\begin{equation}
\varepsilon ^{A,B}\varepsilon ^{C,D}=\varepsilon ^{A,D}\delta
_{BC} ,\hspace{0.5in}\left( \varepsilon ^{A,B}\right) _{CD}=\delta
_{AC}\delta _{BD} ,\label{16}
\end{equation}
where $A,B,C,D=1,2,...,16$. With the help of the matrices
$\varepsilon ^{A,B}$ and Eq. (15), equations (14) become
\[
\biggl \{\partial _\nu \biggl [\varepsilon ^{\mu ,[\mu \nu
]}+\varepsilon ^{[\mu \nu ],\mu }+\varepsilon ^{\nu
,0}+\varepsilon ^{0,\nu }+\varepsilon ^{ \widetilde{\nu
},\widetilde{0}}+\varepsilon ^{\widetilde{0},\widetilde{\nu } }+
\]
\begin{equation}
+\frac 12 e_{\mu \nu \rho \omega }\left( \varepsilon
^{\widetilde{\mu },[\rho \omega ]}+\varepsilon ^{[\rho \omega
],\widetilde{\mu }}\right) \biggr ] _{AB}+ \label{17}
\end{equation}
\[
+\left[ m\left( \varepsilon ^{\mu ,\mu }+\varepsilon
^{\widetilde{\mu }, \widetilde{\mu }}+\frac 12\varepsilon ^{[\mu
\nu ],[\mu \nu ]}\right) +\frac{m_0^2}{m}\left( \varepsilon
^{0,0}+\varepsilon ^{\widetilde{0},\widetilde{0}}\right) \right]
_{AB}\biggr \}\Psi _B(x)=0 .
\]
We note that the matrices
\begin{equation}
P_1=\varepsilon ^{\mu ,\mu }+\varepsilon ^{\widetilde{\mu
}\widetilde{\mu }}+\frac 12\varepsilon ^{[\mu \nu ],[\mu \nu ]} ,
,\hspace{0.5in}P_0=\varepsilon ^{0,0}+\varepsilon
^{\widetilde{0},\widetilde{0}} \label{18}
\end{equation}
are the projection matrices which obey the equations
\[
P_0 P_1= P_1 P_0=0 ,\hspace{0.5in} P_1+P_0=I_{16} ,
\]
where the $I_{16}$ is the unit $16\times 16-$matrix. Introducing
the matrix
\begin{equation}
\Gamma _\nu =\varepsilon ^{\mu ,[\mu \nu ]}+\varepsilon ^{[\mu \nu
],\mu }+\varepsilon ^{\nu ,0}+\varepsilon ^{o,\nu }+\varepsilon
^{\widetilde{\nu }, \widetilde{0}}+\varepsilon
^{\widetilde{0},\widetilde{\nu }}+\frac 12e_{\mu \nu \rho \omega
}\left( \varepsilon ^{\widetilde{\mu },[\rho \omega ]}+\varepsilon
^{[\rho \omega ],\widetilde{\mu }}\right) , \label{19}
\end{equation}
Eq. (16) becomes
\begin{equation}
\left( \Gamma _\nu \partial _\nu +mP_1+\frac{m_0^2}{m}P_0\right)
\Psi (x)=0 . \label{20}
\end{equation}
Eq. (20) is the matrix relativistic wave equation which describes
the system of the vector (pseudovector) and scalar (pseudoscalar)
fields with the masses $m$ and $m_0$, respectively. The $16\times
16-$matrix $\Gamma _\nu $ can be cast as follows:
\[
\Gamma _\nu =\beta _\nu ^{(+)}+\beta _\nu ^{(-)} ,
\hspace{0.3in}\beta _\nu ^{(+)}=\beta _\nu ^{(1)}+\beta _\nu
^{(\widetilde{0})} ,\hspace{0.3in}\beta _\nu ^{(-)}=\beta _\nu
^{(\widetilde{1})}+\beta _\nu ^{(0)} ,
\]
\begin{equation}
\beta _\nu ^{(1)}=\varepsilon ^{\mu ,[\mu \nu ]}+\varepsilon
^{[\mu \nu ],\mu } ,\hspace{0.3in}\beta _\nu
^{(\widetilde{1})}=\frac 12e_{\mu \nu \rho \omega }\left(
\varepsilon ^{\widetilde{\mu },[\rho \omega ]}+\varepsilon ^{[\rho
\omega ],\widetilde{\mu }}\right) , \label{21}
\end{equation}
\[
\beta _\nu ^{(\widetilde{0})}=\varepsilon ^{\widetilde{\nu
},\widetilde{0} }+\varepsilon ^{\widetilde{0},\widetilde{\nu }} ,
\hspace{0.3in}\beta _\nu ^{(0)}=\varepsilon ^{\nu ,0}+\varepsilon
^{0,\nu } .
\]
The 10-dimensional matrices $\beta _\nu ^{(1)}$, $\beta _\nu
^{(\widetilde{1})}$ and 5-dimensional matrices $\beta _\nu
^{(0)}$, $\beta _\nu ^{(\widetilde{0})}$ obey the algebra
\cite{Petiau}, \cite{Duffin}, \cite{Kemmer}:
\begin{equation}
\beta _\mu\beta _\nu \beta _\alpha +\beta _\alpha \beta _\nu \beta
_\mu =\delta _{\mu \nu }\beta _\alpha +\delta _{\alpha \nu }\beta
_\mu  .\label{22}
\end{equation}
The 16-dimensional matrices $\beta _\nu ^{(+)}$, $\beta _\nu
^{(-)}$ realize the reducible representations of the
Petiau-Duffin-Kemmer algebra (22) \cite{Borgardt}. The 16$\times$
16-matrix $\Gamma _\nu $ satisfies the Dirac algebra:
\begin{equation}
\Gamma _\nu \Gamma _\mu +\Gamma _\mu \Gamma _\nu =2\delta _{\mu
\nu } .\label{23}
\end{equation}
In the case when the masses of all states are the same, $m_0=m$,
Eq. (23) is transformed into the matrix form of the the
Dirac-K\"ahler equations \cite{Kruglov1}:
\begin{equation}
\left( \Gamma _\nu \partial _\nu +m\right) \Psi (x)=0 .\label{24}
\end{equation}

\section{Matrix form of 8-component equations}

To get the matrix form of 8-component equations (12), we introduce
four-component wave functions:
\begin{equation}
\xi (x)=-im\left(
\begin{array}{c}
M_a(x) \\
M_4(x)
\end{array}
\right) ,\hspace{0.3in}\chi (x)=\left(
\begin{array}{c}
\widetilde{M}_a(x) \\
M(x)
\end{array}
\right) , \label{25}
\end{equation}
where $\widetilde{M}_a(x)=(1/2)\epsilon _{amn}M_{mn}(x)$. With the
help of Eqs. (25), the system of P-noninvariant wave equations
(12), may be cast in the form
\[
\alpha _\mu \partial _\mu \xi (x)=m\chi (x),
\]
\vspace{-8mm}
\begin{equation}
\label{26}
\end{equation}
\vspace{-8mm}
\[
\overline{\alpha }_\mu \partial _\mu \chi (x)=m\xi (x),
\]
where we introduce matrices:
\[
\alpha _1=\left(
\begin{array}{cccc}
0 & 0 & 0 & -i \\
0 & 0 & -i & 0 \\
0 & i & 0 & 0 \\
ia & 0 & 0 & 0
\end{array}
\right) ,\hspace{0.3in}\alpha _2=\left(
\begin{array}{cccc}
0 & 0 & i & 0 \\
0 & 0 & 0 & -i \\
-i & 0 & 0 & 0 \\
0 & ia & 0 & 0
\end{array}
\right) ,
\]
\vspace{-7mm}
\begin{equation}
\label{27}
\end{equation}
\vspace{-7mm}
\[
\alpha _3=\left(
\begin{array}{cccc}
0 & -i & 0 & 0 \\
i & 0 & 0 & 0 \\
0 & 0 & 0 & -i \\
0 & 0 & ia & 0
\end{array}
\right) ,\hspace{0.3in}\alpha_4=\left(
\begin{array}{cccc}
i & 0 & 0 & 0 \\
0 & i & 0 & 0 \\
0 & 0 & i & 0 \\
0 & 0 & 0 & ia
\end{array}
\right) ,
\]
\[
\overline{\alpha } _1=\left(
\begin{array}{cccc}
0 & 0 & 0 & -i \\
0 & 0 & -i & 0 \\
0 & i & 0 & 0 \\
i & 0 & 0 & 0
\end{array}
\right) ,\hspace{0.3in}\overline{\alpha } _2=\left(
\begin{array}{cccc}
0 & 0 & i & 0 \\
0 & 0 & 0 & -i \\
-i & 0 & 0 & 0 \\
0 & i & 0 & 0
\end{array}
\right) ,
\]
\vspace{-7mm}
\begin{equation}
\label{28}
\end{equation}
\vspace{-7mm}
\[
\overline{\alpha } _3=\left(
\begin{array}{cccc}
0 & -i & 0 & 0 \\
i & 0 & 0 & 0 \\
0 & 0 & 0 & -i \\
0 & 0 & i & 0
\end{array}
\right) ,\hspace{0.3in}\overline{\alpha }_4=\left(
\begin{array}{cccc}
-i & 0 & 0 & 0 \\
0 & -i & 0 & 0 \\
0 & 0 & -i & 0 \\
0 & 0 & 0 & -i
\end{array}
\right) .
\]
It is easy to verify that four-dimensional matrices
$\overline{\alpha }_a$ ($a=1,2,3$) obeys the Pauli commutation
relations:
\begin{equation}
\left\{ \overline{\alpha } _i,\overline{\alpha } _k\right\}
=2\delta _{ik} ,\hspace{0.3in} \left[ \overline{\alpha }
_i,\overline{\alpha } _k\right] =2i\epsilon _{ikl}\overline{\alpha
} _l ,\label{29}
\end{equation}
where $\epsilon _{123}=1$. Note, that matrices $\overline{\alpha
}_a$ may be obtained from matrices $\alpha _a$ by setting $a=1$,
and $\overline{\alpha }_4=-iI_4$. Equations (26) can be also
written in the form of one 8-component equation:
\begin{equation}
\beta _\mu \partial _\mu \varphi (x)+m\varphi (x)=0 ,  \label{30}
\end{equation}
where the matrices of the equation $\beta _\mu$ and 8-component
wave function $\varphi (x)$ are given by
\begin{equation}
\varphi (x)=\left(
\begin{array}{c}
\chi (x) \\
\xi (x)
\end{array}
\right) ,\hspace{0.3in}\beta _\mu =-\left(
\begin{array}{cc}
0 & \alpha _\mu \\
\overline{\alpha }_\mu & 0
\end{array}
\right) . \label{31}
\end{equation}
In the case $a=1$ when the masses scalar and vector states equal
each other, $m=m_0$, matrices $\beta _\mu$ (at $a=1$) obey the
Dirac commutation relations (23) (see \cite{monogr},
\cite{Kruglov6}).

In the same manner, to obtain the matrix form of 8-component
equations (13), we introduce four-component wave functions:
\begin{equation}
\xi ^{\prime }(x)=-im\left(
\begin{array}{c}
N_a(x) \\
N_4(x)
\end{array}
\right) ,\hspace{0.3in}\chi ^{\prime }(x)=\left(
\begin{array}{c}
\widetilde{N}_a(x) \\
N(x)
\end{array}
\right) , \label{32}
\end{equation}
where $\widetilde{N} _a(x)=(1/2)\epsilon _{amn}N_{mn}(x)$. Then
Eqs. (13) can be represented by
\[
\alpha _\mu ^{\prime }\partial _\mu \xi ^{\prime }(x)=m\chi
^{\prime }(x),
\]
\vspace{-7mm}
\begin{equation}
\label{33}
\end{equation}
\vspace{-7mm}
\[
\overline{\alpha }_\mu ^{\prime }\partial _\mu \chi ^{\prime
}(x)=m\xi ^{\prime }(x),
\]
where
\[
\alpha _1^{\prime }=\left(
\begin{array}{cccc}
0 & 0 & 0 & i \\
0 & 0 & -i & 0 \\
0 & i & 0 & 0 \\
ia & 0 & 0 & 0
\end{array}
\right) ,\hspace{0.3in}
 \alpha _2^{\prime }=\left(
\begin{array}{cccc}
0 & 0 & i & 0 \\
0 & 0 & 0 & i \\
-i & 0 & 0 & 0 \\
0 & ia & 0 & 0
\end{array}
\right) ,
\]
\[ \alpha _3^{\prime }=\left(
\begin{array}{cccc}
0 & -i & 0 & 0 \\
i & 0 & 0 & 0 \\
0 & 0 & 0 & i \\
0 & 0 & ia & 0
\end{array}
\right) ,\hspace{0.3in}
 \alpha _4^{\prime }=\left(
\begin{array}{cccc}
-i & 0 & 0 & 0 \\
0 & -i & 0 & 0 \\
0 & 0 & -i & 0 \\
0 & 0 & 0 & ia
\end{array}
\right) ,
\]
\vspace{-7mm}
\begin{equation}
\label{34}
\end{equation}
\vspace{-7mm}
\[
\overline{\alpha }_1^{\prime }=\left(
\begin{array}{cccc}
0 & 0 & 0 & -i \\
0 & 0 & -i & 0 \\
0 & i & 0 & 0 \\
-i & 0 & 0 & 0
\end{array}
\right) ,\hspace{0.3in}
\overline{\alpha }_2^{\prime }=\left(
\begin{array}{cccc}
0 & 0 & i & 0 \\
0 & 0 & 0 & -i \\
-i & 0 & 0 & 0 \\
0 & -i & 0 & 0
\end{array}
\right) ,
\]
\[
\overline{\alpha }_3^{\prime }=\left(
\begin{array}{cccc}
0 & -i & 0 & 0 \\
i & 0 & 0 & 0 \\
0 & 0 & 0 & -i \\
0 & 0 & -i & 0
\end{array}
\right) , \hspace{0.3in}
\overline{\alpha }_4^{\prime }=\left(
\begin{array}{cccc}
i & 0 & 0 & 0 \\
0 & i & 0 & 0 \\
0 & 0 & i & 0 \\
0 & 0 & 0 & -i
\end{array}
\right) .
\]
Two equations (33) are rewritten by one 8-component matrix
equation
\begin{equation}
\beta _\mu ^{\prime }\partial _\mu \varphi ^{\prime }(x)+m\varphi
^{\prime }(x)=0 ,  \label{35}
\end{equation}
where matrices $\beta _\mu ^{\prime }$ and columns $\varphi
^{\prime }(x)$ are
\begin{equation}
\varphi ^{\prime }(x)=\left(
\begin{array}{c}
\chi ^{\prime }(x) \\
\xi ^{\prime }(x)
\end{array}
\right) ,\hspace{0.3in}\beta _\mu ^{\prime }=-\left(
\begin{array}{cc}
0 & \alpha _\mu ^{\prime } \\
\overline{\alpha }_\mu ^{\prime } & 0
\end{array}
\right) . \label{36}
\end{equation}
The matrices $\beta _\mu ^{\prime }$ also at $a=1$ obey the Dirac
algebra (23) (see \cite{monogr}, \cite{Kruglov6}. One may combine
Eqs. (30), (35) in the 16-component relativistic wave equation, as
follows
\begin{equation}
\left( \Pi _\mu \partial _\mu +m\right) \Psi (x)=0 ,  \label{37}
\end{equation}
where
\begin{equation}
\Psi (x)=\left(
\begin{array}{c}
\varphi (x) \\
\varphi ^{\prime }(x)
\end{array}
\right) ,\hspace{0.3in}\Pi _\mu =\left(
\begin{array}{cc}
\beta _\mu & 0 \\
0 & \beta _\mu ^{\prime }
\end{array}
\right) .  \label{38}
\end{equation}
The 16$\times$16 - matrices $\Pi_\mu $ obey at $a=1$ the Dirac
algebra (23). The whole system of equations (12), (13) is
equivalent to Eq. (37) or Eq. (20).  Equation (12) (and (13)) as
well as Eq. (30) (and (35)) are parity noninvariant separately and
at the same time the system of the two equations (12), (13) (or
Eq. (37)) are P-invariant.

We note that at the particular case $m=m_0$ ($a=1$), Eq. (30) and
Eq. (35) are invariant under the $GL(2,c)$ symmetry group
\cite{Kruglov6}, \cite{monogr}. In this case Eq. (20) and Eq. (37)
are equivalent to the Dirac-K\"{a}hler equation and possess the
$GL(4,c)$ internal symmetry group, but the Lagrangian is invariant
under the $SO(4,2)$ group \cite{Kruglov1}, \cite{monogr}. The
condition $m\neq m_0$ ($a\neq1$) eliminates the degeneracy and
destroys the symmetry.

\section{The Lorentz covariance and symmetry of
16-component wave equation}

At the Lorentz transformations of the coordinates:
\begin{equation}
x_\mu ^{\prime }=L_{\mu \nu }x_\nu ^{\prime } , \label{39}
\end{equation}
the matrix $L=\{L_{\mu \nu }\}$ obeys the equation
\begin{equation}
L_{\mu \alpha }L_{\nu \alpha }=\delta _{\mu \nu } . \label{40}
\end{equation}
The wave function (15) under the Lorentz coordinates
transformations (39) is converted into
\begin{equation}
\Psi ^{\prime }(x^{\prime })=T\Psi (x) , \label{41}
\end{equation}
where the $T$ is the 16-dimensional reducible tensor
representation of the Lorentz group. Taking into account that at
the Lorentz transformations (39) the derivatives $\partial _\mu $
transforms as $\partial _\mu ^{\prime }=L_{\mu \nu }\partial _\nu
$, the first order wave equation (20) becomes
\[
\left( \Gamma _\nu \partial _\nu^{\prime }
+mP_1+\frac{m_0^2}{m}P_0\right) \Psi^{\prime } (x^{\prime
})
\]
\vspace{-7mm}
\begin{equation}
\label{42}
\end{equation}
\vspace{-7mm}
\[
=\left( \Gamma _\mu L_{\mu\nu}\partial _\mu
 +mP_1+\frac{m_0^2}{m}P_0\right)T\Psi (x)=0 .
\]
Eq. (20) is form-invariant if the equalities
\begin{equation}
\Gamma _\mu TL_{\mu \nu }=T\Gamma _\nu , \hspace{0.3in}P_1 T=TP_1
, \hspace{0.3in}P_0 T=TP_0 \label{43}
\end{equation}
are satisfied. Let us use the infinitesimal Lorentz
transformations (39) with the matrix
\begin{equation}
L_{\mu \nu }=\delta _{\mu \nu }+\varepsilon _{\mu \nu } ,
\hspace{0.5in} \varepsilon _{\mu \nu }=-\varepsilon _{\nu \mu } ,
\label{44}
\end{equation}
and the infinitesimal transformations (41) with the matrix
\begin{equation}
T=1+\frac 12\varepsilon _{\mu \nu }J_{\mu \nu } , \label{45}
\end{equation}
where $1=I_{16}$, $J_{\mu \nu }$ are the generators of the Lorentz
group in $16-$dimensional space, $\varepsilon _{\mu \nu }$ are six
parameters defining rotations and boosts. Replacing equations
(44), (45) into Eqs. (43) and taking into consideration the
smallness of parameters $\varepsilon _{\mu \nu }$, we arrive at
the equations
\begin{equation}
\Gamma _\mu J_{\alpha \nu }-J_{\alpha \nu }\Gamma _\mu =\delta
_{\alpha \mu }\Gamma _\nu -\delta _{\nu \mu }\Gamma _\alpha
,\hspace{0.3in} P_1 J_{\mu\nu}=J_{\mu\nu}P_1 , \hspace{0.3in}P_0
J_{\mu\nu}=J_{\mu\nu}P_0 \label{46}
\end{equation}
It is easy to verify with the help of formulas of Appendix A, that
generators
\begin{equation}
J_{\mu \nu }=\frac 14\left( \Gamma _{[\mu} \Gamma _{\nu ]}
+\overline{\Gamma }_{[\mu} \overline{\Gamma }_{\nu ]}\right) ,
\label{47}
\end{equation}
satisfy Eqs. (46). The matrices $\overline{\Gamma }_\nu $ are
given by \cite{Borgardt}
\begin{equation}
\overline{\Gamma} _\nu =\beta _\nu ^{(+)}-\beta _\nu ^{(-)} ,
  \label{48}
\end{equation}
and also obey the Dirac algebra (23) and commute with the matrices
$\Gamma _\mu $:
\[
\Gamma _\mu \overline{\Gamma }_\nu=\overline{\Gamma }_\nu \Gamma
_\mu .
\]
Generators of the Lorentz representation (47) are also
valid in the case of $m=m_0$ \cite{Kruglov1}, \cite{monogr}. The
16-dimensional matrix of the finite transformations (41) is given
by
\begin{equation}
T=\exp \left( \frac 12\varepsilon _{\mu \nu }J_{\mu \nu }\right) .
\label{49}
\end{equation}

Let us consider a relativistically invariant bilinear form
\begin{equation}
\overline{\Psi }(x)\Psi (x)=\Psi ^{+}(x)\eta \Psi (x) , \label{50}
\end{equation}
where $\Psi ^{+}$ is the Hermitian-conjugate wave function and
$\eta$ is the Hermitianizing matrix in 16-dimensional space. As
for the case of Dirac-K\"{a}hler equation in the matrix form
\cite{Borgardt}, \cite{Kruglov1}, \cite{monogr}, it is given by:
\begin{equation}
\eta =\Gamma _4\overline{\Gamma }_4 . \label{51}
\end{equation}
The matrix $\eta$ obeys the necessary (see, for example
\cite{Bogush}, \cite{Fedorov}) equations
\[
\eta \Gamma _i=-\Gamma _i\eta \hspace{0.3in}(i=1,2,3) ,
\hspace{0.3in}\eta \Gamma _4=\Gamma _4\eta .
\]

Now we consider the internal symmetry group of Eq. (20). With the
help of Eqs. (16), (18), (21), (48) it is easy to show that the
matrix
 $\overline{\Gamma }_5=\overline{\Gamma
}_1\overline{\Gamma }_2 \overline{\Gamma }_3\overline{\Gamma }_4 $
commutes with the operator of Eq. (20):
\begin{equation}
\overline{\Gamma }_5\left( \Gamma _\mu
\partial_\mu +mP_1 +\frac{m_0^2}{m}P_0\right)=\left( \Gamma _\mu
\partial_\mu +mP_1 +\frac{m_0^2}{m}P_0\right)\overline{\Gamma }_5
. \label{52}
\end{equation}
This means that the matrix $\overline{\Gamma }_5$ can be
considered as the generator of the symmetry group $GL(1,c)$ of Eq.
(20):
\begin{equation}
\Psi ^{\prime }(x)=\exp \left(\omega\overline{\Gamma }_5\right)
\Psi (x) ,
 \label{53}
\end{equation}
where $\omega$ is the complex group parameter. The Lagrangian
corresponding to the first order relativistic wave equation (20)
is given by
\begin{equation}
\mathcal{L}=-\frac 12\left[ \overline{\Psi }(x)\left( \Gamma _\mu
\partial_\mu +mP_1 +\frac{m_0^2}{m}P_0\right) \Psi (x)\right] . \label{54}
\end{equation}
The Lagrangian (54) is invariant under the transformations (53) at
the restriction: $\omega^{*}=\omega $, i.e $\omega $ is a real
parameter. This constrain leaves $GL(1,R)$ subgroup of the
Lagrangian symmetry. The transformations of the Lorentz group (49)
commute with transformations (53) of the internal symmetry group.
As a result, the parameter of this group is a scalar.

According to Noether's theorem, the invariance of the Lagrangian
(54) under the transformation (53) provides the conservation
tensor (see also \cite{Kruglov1}, \cite{monogr} for the case
$m=m_0$):
\begin{equation}
 K_\mu = \overline{\Psi }(x)\Gamma _\mu
\overline{\Gamma }_5\Psi (x) ,\hspace{0.3in}\partial_\mu K_\mu =0
. \label{55}
\end{equation}
The subgroup $U(1)$ of gauge transformations: $\Psi ^{\prime
}(x)=\exp \left( i\alpha \right) \Psi (x)$ ($\alpha ^{*}=\alpha $)
leads to the conservation of four-current $J_\mu =i\overline{\Psi
}(x)\Gamma _\mu \Psi (x)$: $\partial_\mu J_\mu =0$. In the case of
the degeneracy $m=m_0$, the symmetry group of the Lagrangian (54)
is the $SO(4,2)$ group \cite{Kruglov1}, \cite{monogr}.

Using the formulas of Appendix A, we obtain the matrix
$\overline{\Gamma }_5=\overline{\Gamma }_1\overline{\Gamma
}_2\overline{\Gamma }_3\overline{\Gamma }_4$:
\begin{equation}
\overline{\Gamma }_5=-\varepsilon ^{\widetilde{0},0}-\varepsilon
^{0,\widetilde{0}}-\varepsilon ^{\widetilde{\mu },\mu
}-\varepsilon ^{{\mu },\widetilde{\mu}}-\frac 12 e_{\mu \nu \rho
\omega }\varepsilon ^{[\mu \nu],[\rho \omega ]} . \label{56}
\end{equation}
With the aid of Eqs. (15), (56) the infinitesimal transformation
(53): $\delta \Psi (x)= \overline{\Gamma }_5\Psi (x)\delta \omega$
leads to the transformations of the fields (15):
\[
\delta \psi_0 (x)=-\widetilde{\psi}_0 (x)\delta\omega
,\hspace{0.3in}\delta \widetilde{\psi}_0 (x)=-\psi_0
(x)\delta\omega , \hspace{0.3in} \delta \psi_\mu
(x)=-\widetilde{\psi}_\mu (x)\delta\omega ,
\]
\vspace{-7mm}
\begin{equation}
\label{57}
\end{equation}
\vspace{-7mm}
\[
\delta \widetilde{\psi}_\mu (x)=-\psi_\mu (x)\delta\omega
,\hspace{0.3in} \delta
\psi_{[\mu\nu]}(x)=-\frac{1}{2}e_{\mu\nu\varrho\sigma}
\psi_{[\varrho\sigma]}(x)\delta\omega .
\]
The finite transformation (53) can be represented as
\[
 \Psi'(x)=\left(\cosh \omega + \overline{\Gamma }_5
 \sinh \omega \right)\Psi (x) ,
\]
which is an analog of dual transformations in electrodynamics but
with the imaginary value of the angle $\omega$ \cite{monogr}.

\section{Quaternion form of field equations}

Now we obtain the quaternion form of equations (1)-(3).
Quaternions may be considered as a generalization (doubling) of
the complex numbers (see, for instance, \cite{Casanova}).
Quaternions are useful tools for analyzing the symmetry of fields.

The quaternion algebra consists of four basis elements $ e_\mu
=(e_k,e_4)$ (see, for example [24]) with the products:
\[
e_4^2=1,~~e_1^2=e_2^2=e_3^2=-1 ,~~e_1e_2=-e_2e_1=e_3
,~~e_2e_3=-e_3e_2=e_1 ,
\]
\vspace{-7mm}
\begin{equation}
\label{58}
\end{equation}
\vspace{-7mm}
\[
e_3e_1=-e_1e_3=e_2 ,~~ e_4e_m=e_me_4=e_m ~~(m=1, 2, 3) ,
\]
where $e_4$ is the unit element. The biquaternion (or complex
quaternion) $q$ is defined as
\begin{equation}
q=q_\mu e_\mu =q_me_m+q_4e_4 ,  \label{59}
\end{equation}
where $q_\mu $ are complex numbers. With the help of Eqs. (58),
the product of two quaternions, $q$, $q^{\prime }$, is given by:
\begin{equation}
qq^{\prime }=\left( q_4q_4^{\prime }-q_mq_m^{\prime }\right)
e_4+\left( q_4^{\prime }q_m+q_4q_m^{\prime }+\epsilon
_{mnk}q_nq_k^{\prime }\right) e_m . \label{60}
\end{equation}
It should be noted that the combined law for three quaternions:
$\left( q_1q_2\right) q_3=q_1\left( q_2q_3\right)$ holds. The
quaternion conjugation (hyperconjugation) is defined as
\begin{equation}
\overline{q}=q_4e_4-q_me_m\equiv q_4-\mathbf{q} ,  \label{61}
\end{equation}
with the equalities
$\overline{q_1+q_2}=\overline{q}_1+\overline{q}_2$, $
\overline{q_1q_2}=\overline{q}_2\overline{q}_1$.

To get the quaternion form of Eqs. (1), we rewrite them as
follows:
\[
\partial _0 E_m(x)+\partial _m \varphi (x)-\epsilon_{mnk}\partial_n H_k (x)
 -m^2\varphi _m (x)=0 ,
\]
\[
\partial _m E_m(x)+\partial _0 \varphi (x)+m^2\varphi (x)=0 ,
\]
\vspace{-7mm}
\begin{equation}
\label{62}
\end{equation}
\vspace{-7mm}
\[
\partial _0 H_m(x)+\partial _m \widetilde{\varphi}(x)+\epsilon_{mnk}
\partial_n E_k (x)-m^2\widetilde{\varphi} _m (x)=0 ,
\]
\[
\partial _m H_m(x)+\partial _0 \widetilde{\varphi}
(x)+m^2\widetilde{\varphi}(x)=0 ,
\]
where $\varphi _\mu (x)=(\varphi _m (x), i\varphi _0 (x))$, $H_m
(x)=\frac 12\epsilon _{mnk}\varphi _{nk}(x)$, $E_m (x)=i\varphi
_{m4}(x)$, $\partial _0=\partial _t$ ($t$ is the time).
Introducing the biquaternions
\[
\nabla =e_\mu \partial _\mu ,\hspace{0.3in}F=F_\mu e_\mu
,\hspace{0.3in} G=G_\mu e_\mu ,
\]
\vspace{-7mm}
\begin{equation}
\label{63}
\end{equation}
\vspace{-7mm}
\[
F_m=H_m (x)-iE_m (x) ,~~F_4=-\varphi (x)-i\widetilde{\varphi}
(x),~~G_\mu =\varphi_\mu (x)+i\widetilde{\varphi}_\mu (x) ,
\]
with the help of Eqs. (60), we represent Eqs. (62) as follows
\begin{equation}
\nabla F+m^2G=0 .   \label{64}
\end{equation}
So, quaternion equation (64) is equivalent to Eqs. (1). To
represent Eqs. (2), (3) in quaternion form, one may cast them into
\[
\varphi (x)=a\left(\partial _0 \varphi_0 (x)+\partial_m \varphi _m
(x)\right) ,~~\widetilde{\varphi} (x)=a\left(\partial _0
\widetilde{\varphi}_0 (x)+\partial_m \widetilde{\varphi} _m
(x)\right) ,
\]
\begin{equation}
H_m(x)=-\partial _m \widetilde{\varphi}_0 (x)-\partial_0
\widetilde{\varphi} _m (x)+\epsilon_{mnk}\partial_n \varphi_k (x)
, \label{65}
\end{equation}
\[
E_m(x)=-\partial _m \varphi_0 (x)-\partial_0 \varphi _m
(x)-\epsilon_{mnk}\partial_n \widetilde{\varphi}_k (x) .
\]

It easy to show with the help of Eq. (60) that equations
\[
\overline{\nabla }G=\partial_\mu G_\mu e_4+\left(\partial_4 G_m
-\partial_m G_4 -\epsilon_{mnk}\partial_n G_k \right)e_m
,~~\overline{\nabla }=\overline{e}_\mu \partial _\mu ,
\]
\[
\overline{G}\overleftarrow{\nabla}=\partial_\mu G_\mu
e_4-\left(\partial_4 G_m -\partial_m G_4 -\epsilon_{mnk}\partial_n
G_k \right)e_m ,~~\overline{e} _\mu =(e_4,-e_m)
\]
are valid. The arrow above the $\nabla$ shows the action of the
differential operator. Then Eqs. (65) take the form
\begin{equation}
F=\frac{1-a}{2}\overline{G}\overleftarrow{\nabla}-\frac{1+a}{2}\overline{\nabla
}G  , \label{66}
\end{equation}
At the case $a=1$ ($m=m_0$), one arrives at the quaternion form
(Eqs. (64), (66)) of Dirac-K\"{a}hler equation \cite{Kruglov1},
\cite{monogr}.

Under the the Lorentz group transformations, the quaternions $G$,
$F$, $ \nabla $ and $\overline{\nabla }$ are converted into (see
\cite{Kruglov1}, \cite{monogr})
\[
G^L=\overline{L}^{*}GL,\hspace{0.3in}F^L=\overline{L}FL,
\]
\vspace{-8mm}
\begin{equation}
\label{67}
\end{equation}
\vspace{-8mm}
\[
\nabla ^L=\overline{L}^{*}\nabla L,\hspace{0.3in}\overline{\nabla
}^L= \overline{L}\overline{\nabla }L^{*} ,
\]
where quaternions $L$ obey the constrain
$L\overline{L}=\overline{L}L=1$. The field equations (64), (66)
are invariant under the Lorentz transformations (67).

From Eqs. (64), (66) one can obtain the quaternion form of some
equations for the vector fields (the Proca equations, Maxwell's
equations and so on).

\section{Spin projection operators and density matrix}

To obtain all independent solutions of the matrix equation (20) in
the form of matrix-dyads, one needs to construct projection
operators \cite{Fedorov} extracting ``pure" states. With the help
of generators of the Lorentz group (47), we find the spin
projection operator (see \cite{Kruglov1}, \cite{monogr}):
\begin{equation}
\sigma _p=-\frac i{2\mid \mathbf{p}\mid }\epsilon
_{abc}p_aJ_{bc}=-\frac i{4\mid \mathbf{p}\mid }\epsilon
_{abc}p_a\left( \Gamma _b\Gamma _c+ \overline{\Gamma
}_b\overline{\Gamma }_c\right)  \label{68}
\end{equation}
This operator obeys the ``minimal" equation as follows:
\begin{equation}
\sigma _p\left( \sigma _p-1\right) \left( \sigma _p+1\right) =0 .
\label{69}
\end{equation}
According to the general method \cite{Fedorov}, one obtains the
projection operators extracting spin projections $\pm 1$ and $0$:
\begin{equation}
\widehat{S}_{(\pm 1)}=\frac 12\sigma _p\left( \sigma _p\pm
1\right) ,\hspace{0.5in}\widehat{S}_{(0)}=1-\sigma _p^2
.\label{70}
\end{equation}
These operators satisfy the ordinary relationships:
$\widehat{S}_{(\pm 1)}^2=\widehat{S}_{(\pm 1)}$, $\widehat{S}
_{(\pm 1)}\widehat{S}_{(0)}=0$,
$\widehat{S}_{(0)}^2=\widehat{S}_{(0)}$. We consider fields with
different spins, one and zero. To separate these states, one has
to construct the squared spin operator. Such an operator is given
by the squared Pauli-Lubanski vector $\sigma ^2$:
\begin{equation}
\sigma ^2=\left( \frac 1{2m}e_{\mu \nu \alpha \beta }p_\nu
J_{\alpha \beta }\right) ^2=\frac 1{m^2}\left( \frac{1}{2}J_{\mu
\nu }^2p^2-J_{\mu \sigma }J_{\nu \sigma }p_\mu p_\nu \right) .
\label{71}
\end{equation}
One can verify with the aid of Eqs. (47) (see Appendix) that this
operator satisfies the matrix equation
\begin{equation}
\sigma ^2\left( \sigma ^2-2\right) =0 . \label{72}
\end{equation}
Eq. (72) shows that eigenvalues of the squared spin operator
$\sigma ^2$ are $s(s+1)=0$ and $s(s+1)=2$; that corresponds to
spins $s=0$ and $s=1$. The projection operators separating these
states are given by
\begin{equation}
S_{(0)}^2=1-\frac{\sigma ^2}2
,\hspace{0.5in}S_{(1)}^2=\frac{\sigma ^2}2 .\label{73}
\end{equation}
Operators (73) possess the properties
\[
S_{(0)}^2S_{(1)}^2=0 ,~~\left( S_{(0)}^2\right) ^2=S_{(0)}^2 ,
~~\left( S_{(1)}^2\right) ^2=S_{(1)}^2 ,~~ S_{(0)}^2+S_{(1)}^2= 1
.
\]
The projection operators $S_{(0)}^2$, $S_{(1)}^2$ acting on the
wave function extract pure states with spin $0$ and $1$,
respectively.

The system of field equations under consideration includes two
fields with spin one (vector $\psi _\mu $, and pseudovector
$\widetilde{\psi }_\mu $, fields), and two fields with spin zero
(scalar $ \psi _0$, pseudoscalar $\widetilde{\psi }_0$, fields).
So, there is a doubling of the spin states of fields, i.e.
degeneracy. It is necessary to introduce additional projection
operators to separate these states and to have pure spin states.
For this purpose, one may explore the internal symmetry of fields
which mixes scalar and pseudoscalar fields as well as vector and
pseudovector fields, and is given by Eq. (53). We introduce the
projection operator as follows:
\begin{equation}
\Lambda _\lambda =\frac 12\left( 1+\lambda \overline{\Gamma
}_5\right) , \label{74}
\end{equation}
where $\lambda =\pm 1$. Acting this operator on the wave function
(15), one obtains two subsystems of equations (12) and (13)
corresponding to additional quantum number $\lambda =\pm 1$. The
operator (74) obeys the relation $\Lambda _\lambda ^2=\Lambda
_\lambda $ required. It is not difficult to check that all
introduced projection operators (70), (73) and (74) commute with
each other and with the operator of the wave equation (20). So,
the product of these operators
\begin{equation}
\Lambda _\lambda\widehat{S}_{(\pm 1), (0)}S_{(1), (0)}^2 ,
\label{75}
\end{equation}
extracts the states with definite spin, spin projection and
additional quantum number $\lambda =\pm 1$ which is connected with
doubling of spin states. It should be noted that these states are
not states with the definite parity because the operator $\Lambda
_\lambda$ mixes states with the opposite parity. The parity
operator is given by $P=\eta_P \Gamma_4 \overline{\Gamma }_4$, and
it does not commute with the operator $\Lambda _\lambda$. To have
the states with definite energy, we consider Eq. (20) in the
momentum space:
\begin{equation}
\left( \varepsilon i\widehat{p} + m P_1+\frac{m_0^2}{m}P_0\right)
\Psi (p)=0 ,\label{76}
\end{equation}
where $\widehat{p}=p_\mu \Gamma _\mu $, $\varepsilon=1$
corresponds to positive energy, and  $\varepsilon=-1$ -- to
negative energy. It easy to verify, using Eqs. (22), (23), that
the projection operator
\[
M^{(1)}_\varepsilon =\frac{i\left(
p^{(1)}+p^{(\widetilde{1})}\right)\left[ i\left( p^{(1)}+
p^{(\widetilde{1})}\right) - \varepsilon m\right]}{2m^2}
\]
\vspace{-8mm}
\begin{equation}  \label{77}
\end{equation}
\vspace{-8mm}
\[
=\frac{i p^{(1)}\left( i p^{(1)} - \varepsilon
m\right)}{2m^2}+\frac{i p^{(\widetilde{1})}\left( i
p^{(\widetilde{1})} - \varepsilon m\right)}{2m^2} ,
\]
where $p^{(1)}=  p_\nu \beta _\nu ^{(1)}$, $p^{(\widetilde{1})}=
p_\nu \beta _\nu ^{(\widetilde{1})}$, obeys the equality
$(M^{(1)}_\varepsilon )^2=M^{(1)}_\varepsilon $, is a solution of
Eq. (76), and corresponds to states with spin one (see
\cite{Kruglov1}, \cite{monogr}). Using Eqs. (21), (56), one
obtains the equalities
\begin{equation}
 \beta _\nu
^{(1)}\overline{\Gamma }_5=\overline{\Gamma }_5 \beta _\nu
^{(\widetilde{1})} ,~~\beta _\nu
^{(\widetilde{1})}\overline{\Gamma }_5=\overline{\Gamma }_5 \beta
_\nu ^{(1)} .
   \label{78}
\end{equation}
It is easy to see with the help of Eqs. (78) that operator (77)
commutes with the operators (74). Using the properties of the
entire algebra (16), and Eqs. (21), (70), (73), one may verify the
relations \cite{Kruglov1}, \cite{monogr}:
\[
\widehat{S}_{(\pm 1)}p^{( 0)}=\widehat{S}_{(\pm
1)}p^{(\widetilde{0})}=0 ,~~S_{(1)}^2 p^{(0)}=S_{(1)}^2
p^{(\widetilde{0})}=0 ,~~S_{(0)}^2 p^{(1)}=S_{(0)}^2
p^{(\widetilde{1})}=0 ,
\]
\[
p^{(1)}p^{(0)}=p^{(0)}p^{(1)}=p^{(\widetilde{1})}p^{(\widetilde{0})}
=p^{(\widetilde{0})}p^{(\widetilde{1})}=0 ,
\]
\vspace{-7mm}
\begin{equation}
\label{79}
\end{equation}
\vspace{-7mm}
\[
 p^{(1)}p^{(\widetilde{1})}=p^{(\widetilde{1})}p^{(1)}
=p^{(0)}p^{(\widetilde{0})}=p^{(\widetilde{0})}p^{(0)}=0 ,
\]
\[
\widehat{S}_{(0)}p^{( 0)}=p^{( 0)}
,~~\widehat{S}_{(0)}p^{(\widetilde{0})}=p^{(\widetilde{0})} ,~~
S_{(1)}^2 p^{(1)}=p^{(1)}
,~~S^2_{(1)}p^{(\widetilde{1})}=p^{(\widetilde{1})} .
\]
It is checked with the help of Eqs. (79) that the operators (77)
commute with the operators (75). As a result, from Eqs. (75),
(77), we obtain projection operators, in the form of matrix-dyads,
extracting pure states with spin one, spin projection $\pm 1$,
$0$, definite energy, and quantum number $\lambda=\pm 1$:
\[
\Delta ^{(1)}=M^{(1)}_\varepsilon S_{(1)}^2 \widehat{ S}_{(\pm
1)}\Lambda _\lambda = \Psi ^{(1)}\cdot \overline{ \Psi }^{(1)} ,
\]
\vspace{-7mm}
\begin{equation}
\label{80}
\end{equation}
\vspace{-7mm}
\[
\Delta _0^{(1)}=M^{(1)}_\varepsilon S_{(1)}^2 \widehat{
S}_{(0)}\Lambda _\lambda =\Psi _0^{(1)}\cdot \overline{\Psi
}_0^{(1)} ,
\]
 The matrix-dyads $\Delta ^{(1)}$, $\Delta
_0^{(1)}$, correspond to the states with spin one, spin
projections $\pm 1$ and $0$, correspondingly. The construction of
matrix-dyads for the states with spin zero requires finding other
solutions of Eq. (76) corresponding to spin zero, and here it is
not considered. The dyad representation (80) can be applied to
quantum electrodynamic calculations of possesses with the presence
of fields under consideration. The method of computing the traces
of 16-dimensional matrix products was developed in
\cite{Kruglov1}, \cite{monogr}.

\section{Canonical quantization of fields}

To apply the canonical quantization \cite{Dirac}, we consider the
fields $\Psi_A (x)$ in the Lagrangian (54) as ``coordinates" and
the variables $\partial_0\Psi_A (x)$ as ``velocities". The momenta
are defined as
\begin{equation}
\pi_A (x)=\frac{\partial {\cal L}(x)}{\partial \left(\partial_0
\Psi_A (x)\right)}=i\left(\overline{\Psi }(x)\Gamma_4 \right)_A=
i\left(\Psi^+ (x)\overline{\Gamma}_4 \right)_A  .\label{81}
\end{equation}
It should be noted that there are no constraints here. With the
help of the Poisson brackets between ``coordinates" $\Psi_A (x)$
and momenta $\pi_A (x)$, from Eq. (81), one obtains
\begin{equation}
\{\Psi_A (\mathbf{x},t),i\left(\Psi^+ (\mathbf{x}^{\prime }
,t)\overline{\Gamma}_4 \right)_B
\}=\delta_{AB}\delta\left(\mathbf{x}- \mathbf{x}^{\prime }\right)
. \label{82}
\end{equation}
Using the equality $\overline{\Gamma}_4^2 =1$, and the transition
to the quantum theory by the substitution
\[
\{\Psi ,\pi \} =-i\left[\Psi,\pi \right] ,
\]
where $\left[\Psi,\pi \right]= \Psi\pi -\pi \Psi$, we arrive at
the quantum commutators
\begin{equation}
\left[ \Psi _A(x),\Psi^+ _B(x^{\prime })\right] _{t=t^{\prime
}}=\left( \overline{\Gamma} _4\right) _{AB}\delta
(\mathbf{x}-\mathbf{x}^{\prime }) ,\label{83}
\end{equation}
where $A$, $B=1$,$2$,...,$16$. The quantization is performed here
in the same manner as in the case $m=m_0$ (a=1) \cite{Kruglov1},
\cite{monogr}. Using Eqs. (15), (19), one finds the commutation
relations as follows:
\[
\left[ \varphi (x),m\varphi _0^{*}(x^{\prime })\right]
_{t=t^{\prime }}=i\delta (\mathbf{x}-\mathbf{x}^{\prime }) ,
\hspace{0.3in}\left[ \widetilde{ \varphi }(x),m\widetilde{\varphi
}_0^{*}(x^{\prime })\right] _{t=t^{\prime }}=-i\delta
(\mathbf{x}-\mathbf{x}^{\prime }) ,
\]
\begin{equation}
\left[ m\widetilde{\varphi }_k(x),H _{m}^{*}(x^{\prime })\right]
_{t=t^{\prime }}=i\delta _{km}\delta
(\mathbf{x}-\mathbf{x}^{\prime }) ,\label{84}
\end{equation}
\[
\left[ m\varphi _k(x),E_{n}^{*}(x^{\prime })\right] _{t=t^{\prime
}}=i\delta _{kn}\delta (\mathbf{x}-\mathbf{x}^{\prime }) ,
\]
where $H _{m}=(1/2)\epsilon_{mnk}\varphi_{nk}$, $E_m
=i\varphi_{m4}$. Expressing the components $\varphi_0 (x)$,
$\widetilde{ \varphi }_0(x)$, $\varphi_{\mu\nu}$ from Eqs.
(1)-(3), replacing them into Eqs. (84), and taking into
consideration the commutators of fields, one arrives at the
relations
\begin{equation}
\left[ \varphi (x),\partial_0 \varphi^{*}(x^{\prime })\right]
_{t=t^{\prime }}=-im\delta (\mathbf{x}-\mathbf{x}^{\prime }) ,
\label{85}
\end{equation}
\begin{equation}
\left[ \widetilde{ \varphi }(x),\partial_0 \widetilde{\varphi
}^{*}(x^{\prime })\right] _{t=t^{\prime }}=im\delta
(\mathbf{x}-\mathbf{x}^{\prime }) , \label{86}
\end{equation}
\begin{equation}
\left[ \varphi_k(x),\partial_0 \varphi _{m}^{*}(x^{\prime
})\right] _{t=t^{\prime }}=\frac{i}{m}\delta _{km}\delta
(\mathbf{x}-\mathbf{x}^{\prime }) ,\label{87}
\end{equation}
\begin{equation}
\left[ \widetilde{\varphi} _k(x),\partial_0
\widetilde{\varphi}_{n}^{*}(x^{\prime })\right] _{t=t^{\prime
}}=-\frac{i}{m}\delta _{kn}\delta (\mathbf{x}-\mathbf{x}^{\prime
}) . \label{88}
\end{equation}
It follows from Eqs. (85)-(88) that vector and pseudoscalar fields
have the positive metric and pseudovector and scalar fields lead
to negative metric because of the additional sign (-)
\cite{monogr}. The density of the Hamiltonian is defined by the
relation ${\cal H}(x)=\pi_A (x) \partial_0 \Psi_A (x)-{\cal
L}(x)$. With the help of Eq. (81), (15), (48), we find
\[
{\cal H}(x)=i\Psi^+ (x)\overline{\Gamma}_4 \partial_0 \Psi (x)
\]
\begin{equation}
=m\biggl [\varphi_m^* (x)\partial_0 E_m (x) - E_m^* (x)\partial_0
\varphi_m (x)+\varphi_0^* (x)\partial_0 \varphi (x)-\varphi^*
(x)\partial_0 \varphi_0 (x) \label{89}
\end{equation}
\[
- \widetilde{\varphi}_0^* (x)\partial_0
\widetilde{\varphi}(x)+\widetilde{\varphi}^* (x)\partial_0
\widetilde{\varphi}_0 (x)- \widetilde{\varphi}_m^* (x)\partial_0
H_m (x) + H_m^* (x)\partial_0 \widetilde{\varphi}_m (x)\biggr ] .
\]
From the expression for four-current $J_\mu
(x)=i\overline{\Psi}(x)\Gamma_\mu \Psi (x)$, one obtains, with the
aid of Eqs. (15), (48), the total electric charge
\[
Q=\int d^3 x\overline{\Psi}(x)\Gamma_4 \Psi (x)=\int d^3 x\Psi^+
(x)\overline{\Gamma}_4 \Psi (x)
\]
\begin{equation}
 =-im\int d^3 x\biggl [\varphi_m^*
(x)E_m (x) - E_m^* (x)\varphi_m (x)+\varphi_0^* (x)\varphi (x)
-\varphi^* (x)\varphi_0 (x)
 \label{90}
\end{equation}
\[
- \widetilde{\varphi}_0^*
(x)\widetilde{\varphi}(x)+\widetilde{\varphi}^* (x)
\widetilde{\varphi}_0 (x)- \widetilde{\varphi}_m^* (x)H_m (x) +
H_m^* (x)\widetilde{\varphi}_m (x)\biggr ] .
\]
The quantizing procedure here is similar to one in quantum
electrodynamics (QED) for the bispinor fields. But the difference
is in the statistics: bispinor fields obey the Fermi-Dirac
statistics, and 16-component fields $\Psi (x)$ under
consideration, are bosonic fields satisfying the Bose-Einstein
statistics. As a result, we have here commutators (83) instead of
anticommutators in QED.

It follows from Eqs. (89), (90) that neither energy nor charge of
the classical fields under consideration have positive values.
Therefore in quantized theory we have to introduce indefinite
metric \cite{monogr}. The total space of states has two subspaces
with positive ($H_p$) and negative ($H_n$) square norms. According
to Eqs. (85)-(88) the vector and pseudoscalar states possess a
positive square norm, and pseudovector and scalar states have a
negative square norms. The total space of states represents the
direct sum of the two subspaces $H_p$ and $H_n$.

\section{Electromagnetic interaction of fields}

The ``minimal" electromagnetic interaction is introduced by the
substitution $\partial_\mu \rightarrow D_\mu=\partial_\mu -ieA_\mu
$, where $A_\mu $ is the vector-potential of the electromagnetic
field. The relativistic non-minimal electromagnetic interaction
may be taken into account by consideration of the operators (see
Appendix):
\begin{equation}
\overline{\Gamma}_{[\mu}\overline{\Gamma}_{\nu]}{\cal F}_{\mu\nu}
,~~ \Gamma_{[\mu}\Gamma_{\nu]}{\cal F}_{\mu\nu}
,~~\overline{\Gamma}_{[\mu}\Gamma_{\nu]}{\cal F}_{\mu\nu} ,
\label{91}
\end{equation}
where ${\cal F}_{\mu\nu}=\partial_\mu A_\nu -\partial_\nu A_\mu $
is the strength of the electromagnetic field. Introducing
phenomenological parameters $\kappa_1$, $\kappa_2$, $\kappa_3$ in
the operators (91), and the replacement $\partial_\mu \rightarrow
D_\mu$, one obtains from Eq. (20) the relativistic wave equation
of the first order for the interaction of bosonic fields under
consideration with the electromagnetic field:
\[
\biggl [ \Gamma _\nu D _\nu +mP_1+\frac{m_0^2}{m}P_0
\]
\vspace{-8mm}
\begin{equation}
\label{92}
\end{equation}
\vspace{-8mm}
\[
+\frac{1}{4}\left(\kappa_1\overline{\Gamma}_{[\mu}
\overline{\Gamma}_{\nu]} +\kappa_2\Gamma_{[\mu}\Gamma_{\nu]}
+\kappa_3\overline{\Gamma}_{[\mu}\Gamma_{\nu]} \right){\cal
F}_{\mu\nu}\biggr] \Psi (x)=0 . \label{20}
\]
To clear up the physical meaning of constants $\kappa_1$,
$\kappa_2$, $\kappa_3$ introduced, we rewrite Eq. (92) in tensor
form. With the help of Eqs. (97)-(99) (see Appendix), (15), (19),
from Eq. (20), one can obtain the system of equations for
interacting scalar, vector, pseudoscalar and pseudovector fields
with electromagnetic field:
\[
D_\mu \psi_\mu +\frac{m_0^2}{m}\psi_0 +\frac{1}{2}\left(\kappa_2
-\kappa_1 -\kappa_3\right){\cal F}_{\mu\nu}\psi_{[\mu\nu]}=0 ,
\]
\[
D_\mu \widetilde{\psi}_\mu+\frac{m_0^2}{m}\widetilde{\psi}_0
+\frac{i}{2}\left(\kappa_2 -\kappa_1 +\kappa_3\right)\widetilde{
{\cal F}}_{\mu\nu}\psi_{[\mu\nu]}=0 ,
\]
\[
D_\nu \psi_{[\mu\nu]} +D_{\mu}\psi_0 + m\psi_\mu  +
 \left(\kappa_2 +\kappa_1 \right){\cal
F}_{\mu\nu}\psi_{\nu}+i\left(\kappa_1 -\kappa_2
-\kappa_3\right)\widetilde{ {\cal
F}}_{\mu\nu}\widetilde{\psi}_{\nu}=0 ,
\]
\vspace{-7mm}
\begin{equation}  \label{93}
\end{equation}
\vspace{-7mm}
\[
iD_\nu \widetilde{\psi}_{[\mu\nu]} +D_{\mu}\widetilde{\psi}_0 +
m\widetilde{\psi}_\mu  +
 \left(\kappa_2 +\kappa_1 \right){\cal
F}_{\mu\nu}\widetilde{\psi}_{\nu}+i\left(\kappa_1 -\kappa_2
+\kappa_3\right)\widetilde{ {\cal F}}_{\mu\nu}\psi_{\nu}=0 ,
\]
\[
D_\nu \psi_{\mu} -D_{\mu}\psi_\nu + e_{\mu\nu\alpha\beta}D_\beta
\widetilde{\psi}_\alpha + m\psi_{[\mu \nu]} +
 \left(\kappa_2 +\kappa_1 \right)\left({\cal
F}_{\nu\beta}\psi_{\mu\beta}- {\cal
F}_{\mu\beta}\psi_{\nu\beta}\right)
\]
\[
+\left(\kappa_1 -\kappa_2 -\kappa_3\right)  {\cal F}_{\mu\nu}
\psi_0 +i\left(\kappa_1 -\kappa_2 +\kappa_3\right)\widetilde{
{\cal F}}_{\mu\nu}\widetilde{\psi}_0=0 ,
\]
where dual tensors are defined by $\widetilde{ {\cal
F}}_{\mu\nu}=(1/2)\varepsilon_{\mu\nu\alpha\beta}{\cal
F}_{\alpha\beta}$, $\widetilde{
\psi}_{[\mu\nu]}=(1/2)\varepsilon_{\mu\nu\alpha\beta}\psi_
{[\alpha\beta]}$ ($\varepsilon_{\mu\nu\alpha\beta}
=(-i)e_{\mu\nu\alpha\beta}$). Replacing the variables in Eq. (93)
with the help of equalities (see Eqs. (1)-(4)) $\psi _0 =-\varphi
$, $\psi _\mu =m\varphi _\mu $, $\psi _{[\mu \nu ]}=\varphi _{\mu
\nu }$, $\widetilde{\psi }_\mu =im\widetilde{ \varphi }_\mu $,
$\widetilde{\psi }_0 =-i\widetilde{\varphi }$, one can express the
scalar and pseudoscalar fields from Eqs. (93) as follows:
\[
\varphi =\frac{m^2}{m_0^2}D_\mu \varphi_\mu
+\frac{m}{2m_0^2}\left(\kappa_2 -\kappa_1 -\kappa_3\right){\cal
F}_{\mu\nu}\varphi_{\mu\nu} ,
\]
\vspace{-7mm}
\begin{equation}  \label{94}
\end{equation}
\vspace{-7mm}
\[
\widetilde{\varphi} =\frac{m^2}{m_0^2}D_\mu
\widetilde{\varphi}_\mu +\frac{m}{2m_0^2}\left(\kappa_2 -\kappa_1
+\kappa_3\right)\widetilde{{\cal F}}_{\mu\nu}\varphi_{\mu\nu} .
\]
Excluding the scalar and pseudoscalar fields in Eqs. (93) with the
aid of Eqs. (94) (with the appropriate replacement of fields), one
obtains the system of interacting vector, pseudovector and tensor
fields:
\[
D_\nu \varphi_{\mu\nu} -\frac{m^2}{m_0^2}D_\mu
D_\alpha\varphi_\alpha -\frac{m}{2m_0^2}\left(\kappa_2 -\kappa_1
-\kappa_3\right)D_\mu\left({\cal
F}_{\alpha\nu}\varphi_{\alpha\nu}\right)
\]
\[
+m^2\varphi_\mu +m\left(\kappa_1 +\kappa_2 \right){\cal
F}_{\mu\nu}\varphi_{\nu}-m\left(\kappa_1 -\kappa_2 -\kappa_3
\right)\widetilde{{\cal F}}_{\mu\nu}\widetilde{\varphi}_{\nu}=0 ,
\]
\[
D_\nu \widetilde{\varphi}_{\mu\nu} -\frac{m^2}{m_0^2}D_\mu
D_\alpha\widetilde{\varphi}_\alpha -\frac{m}{2m_0^2}\left(\kappa_2
-\kappa_1 +\kappa_3\right)D_\mu\left(\widetilde{{\cal
F}}_{\alpha\nu}\varphi_{\alpha\nu}\right)
\]
\begin{equation}
+ m^2 \widetilde{\varphi}_\mu +m\left(\kappa_1 +\kappa_2
\right){\cal F}_{\mu\nu}\widetilde{\varphi}_{\nu}+m\left(\kappa_1
-\kappa_2 +\kappa_3 \right)\widetilde{{\cal
F}}_{\mu\nu}\varphi_{\nu}=0 , \label{95}
\end{equation}
\[
\varphi_{\mu\nu}=D_\mu \varphi_\nu -D_\nu \varphi_\mu
-\varepsilon_{\mu\nu\beta\alpha}D_\beta \widetilde{\varphi}_\alpha
-\frac{\kappa_1 +\kappa_2}{m}\left( {\cal F}_{\nu\beta}\varphi_{
\mu\beta}-{\cal F}_{\mu\beta}\varphi_{ \nu\beta}\right)
\]
\[
+\frac{m}{m_0^2}\left[\left( \kappa_1 -\kappa_2
-\kappa_3\right){\cal F}_{\mu\nu }D_\alpha\varphi_\alpha-\left(
\kappa_1 -\kappa_2 +\kappa_3\right)\widetilde{{\cal F}}_{\mu\nu }
D_\alpha\widetilde{\varphi}_\alpha\right]
\]
\[
+\frac{\left(\kappa_1 -\kappa_2\right)^2
-\kappa_3^2}{2m_0^2}\left(\widetilde{{\cal F}}_{\mu\nu
}\widetilde{{\cal F}}_{\alpha\beta }-{\cal F}_{\mu\nu }{\cal
F}_{\alpha\beta}\right)\varphi_{\alpha\beta} .
\]
It follows from Eqs. (95) that constants $\kappa_1$, $\kappa_2$,
$\kappa_3$ are connected with the anomalous magnetic moment (AMM)
and quadrupole electric moment (KEM) of a particle \cite{Joung}.
The terms in Eqs. (95) containing $m_0$ are contributed from the
scalar and pseudoscalar states. At the limiting case
$m_0\rightarrow\infty$ (masses of a scalar and pseudoscalar states
are infinite), one neglects this contribution because the
corresponding terms approach to zero. At the particular case
$m=m_0$, $\kappa_1 =ie/2m$, $\kappa_2 =\kappa_3 =0$, we arrive at
the description of vector particles with gyromagnetic ratio $g=2$
\cite{Durand}, \cite{monogr}. So, the usage of the representation
of the Lorentz group with high dimension (in our case dimension is
$16$) for the description of vector fields allows us to introduce
more phenomenological constants of electromagnetic interaction.

\section{Conclusion}

Equations describing fields which may exist in two mass ($m$,
$m_0$) and spin (one and zero) states have been considered. Such
fields possess the additional symmetry due to the doubling of spin
states. This symmetry is an analog of dual transformations in
electrodynamics. At the equal masses of spin one and zero fields,
we recover the internal symmetry group $SO(4,2)$ (or locally
isomorphic to the $U(2,2)$ group) investigated in \cite{Kruglov1},
\cite{monogr}. This symmetry allows us to construct gauge theories
with the non-compact groups \cite{monogr}.

The matrices of the relativistic wave equation obtained obey the
Dirac commutation relations. However, the matrix equation contains
the additional projection operators which are connected with two
mass states of scalar and vector fields. At the particular case of
equal masses of scalar and vector states, $m=m_0$, we arrive at
the Dirac-like 16-dimensional equation which is equivalent to the
Dirac-K\"{a}hler equation. Such equation is widely used for
describing quarks (spin 1/2 fields) on the lattice. However, we
consider here bosonic fields which correspond to the Bose-Einstein
statistics. The quantization procedure has been carried out
similar to QED, but with the commutation relations of fields
instead of the anticommutators. Although the quantization of such
fields requires the introduction of the indefinite metric, the
theory of fields under consideration can be considered as an
effective theory.

The density matrices obtained for spin one fields may be applied
for calculations of probabilities of different quantum processes
in a covariant manner. The method considered in \cite{Fedorov}
allows us to make evaluations of physical quantities without using
the matrices of first-order equations.

The considered quaternion form of equations is very convenient for
the study of internal and Lorentz group symmetries. In these
particular cases, one may get from Eqs. (64), (66) well-known and
important Maxwell, Proca, etc. equations in the quaternion form.

We may speculate that the fields under consideration can be used
for theoretical schemes describing different objects including
sub-quark matter. This, however, requires to solve the problem of
the physical interpretation of quantum fields with indefinite
metric.

We have also studied fields describing particles possessing AMM
and KEM and, therefore, having the internal structure. Possibly,
this scheme may be applied for the description of composite
systems in nuclear and particle physics (see \cite{Kirchbach}).

\vspace{5mm} {\bf APPENDIX. Products of 16-Dimensional Matrices}
\vspace{5mm}

Consider products of Petiau-Duffin-Kemmer matrices. It is implied
that all matrices act in 16-dimensional space. Using the
properties of the entire algebra matrices (16), we obtain
\[
\beta _\mu ^{(1)}\beta _\nu ^{(1)}=\varepsilon ^{[\alpha\mu]
,[\alpha \nu]} + \delta_{\mu\nu}\varepsilon^{\alpha,\alpha}
-\varepsilon ^{\nu ,\mu } ,\hspace{0.3in} \beta _\mu
^{(\widetilde{1})}\beta _\nu ^{(1)}=e_{\mu \nu \rho \omega
}\varepsilon ^{\widetilde{\omega },\rho} ,
\]
\[
\beta _\mu ^{(0)}\beta _\nu ^{(1)}=\varepsilon ^{0,[\mu\nu]}
,\hspace{0.3in}\beta _\mu ^{(\widetilde{0})}\beta _\nu
^{(\widetilde{0})}=\delta_{\mu\nu}\varepsilon^{\widetilde{0},\widetilde{0}}
+\varepsilon ^{\widetilde{\mu },\widetilde{\nu }} ,
\]
\begin{equation}
\beta _\mu ^{(\widetilde{1})}\beta _\nu
^{(\widetilde{0})}=\frac{1}{2}e_{\nu\mu \rho \omega }\varepsilon
^{ [\rho\omega],\widetilde{0}} ,\hspace{0.3in}\beta _\mu
^{(1)}\beta _\nu ^{(\widetilde{1})}=e_{\mu\nu\rho \omega
}\varepsilon ^{\omega,\widetilde{\rho}} ,\label{96}
\end{equation}
\[
\beta _\mu ^{(\widetilde{0})}\beta _\nu
^{(\widetilde{1})}=\frac{1}{2}e_{ \mu\nu \rho \omega }\varepsilon
^{\widetilde{0},[\rho\omega]} ,~~\beta _\mu
^{(\widetilde{1})}\beta _\nu ^{(\widetilde{1})}= \delta_{\mu\nu}
\left(\varepsilon^{\widetilde{\alpha},\widetilde{\alpha}} +
\frac{1}{2}\varepsilon ^{[\rho\omega],[\rho\omega]} \right)-
\varepsilon^{\widetilde{\nu},\widetilde{\mu}} + \varepsilon
^{[\alpha\nu],[\mu\alpha]} ,
\]
\[
\beta _\mu ^{(1)}\beta _\nu ^{(0)}=\varepsilon ^{[\nu\mu] ,0}
,\hspace{0.3in}\beta _\mu ^{(0)}\beta _\nu ^{(0)}=\delta_{\mu\nu}
\varepsilon^{0,0} + \varepsilon ^{\mu,\nu} .
\]
We write out only nonzero elements of matrices in 16-dimensional
space. With the help of Eqs. (16), (21), (48), one finds the
antisymmetric product of 16-dimensional Dirac matrices
\[
\frac{1}{2}\Gamma_{[\mu} \Gamma _{\nu
]}\equiv\frac{1}{2}\left(\Gamma_\mu \Gamma _\nu -\Gamma_\nu \Gamma
_\mu \right)=\varepsilon ^{[\alpha\mu],[\alpha\nu]}- \varepsilon
^{[\alpha\nu][ \alpha\mu]} +\varepsilon ^{\mu,\nu}-\varepsilon
^{\nu,\mu} + \varepsilon^{\widetilde{\mu},\widetilde{\nu}} -
\varepsilon^{\widetilde{\nu},\widetilde{\mu}}
\]
\vspace{-8mm}
\begin{equation}  \label{97}
\end{equation}
\vspace{-8mm}
\[
+e_{\mu \nu \rho \omega }\left(\varepsilon ^{\widetilde{\omega
},\rho} - \varepsilon ^{\rho,\widetilde{\omega }}\right) +
\varepsilon ^{0,[\mu\nu]}-\varepsilon ^{[\mu\nu],0} +
\frac{1}{2}e_{\mu\nu\rho\omega}\left(\varepsilon^{\widetilde{0},
[\rho\omega]} -\varepsilon^{[\rho\omega],\widetilde{0}}\right) .
\]
\[
\frac{1}{2}\overline{\Gamma }_{[\mu} \overline{\Gamma }_{\nu
]}\equiv\frac{1}{2}\left(\overline{\Gamma }_\mu \overline{\Gamma
}_\nu -\overline{\Gamma }_\nu \overline{\Gamma
}_\mu\right)=\varepsilon ^{[\alpha\mu],[\alpha\nu]}- \varepsilon
^{[\alpha\nu][ \alpha\mu]} +\varepsilon ^{\mu,\nu}-\varepsilon
^{\nu,\mu} + \varepsilon^{\widetilde{\mu},\widetilde{\nu}} -
\varepsilon^{\widetilde{\nu},\widetilde{\mu}}
\]
\vspace{-8mm}
\begin{equation}  \label{98}
\end{equation}
\vspace{-8mm}
\[
-e_{\mu \nu \rho \omega }\left(\varepsilon ^{\widetilde{\omega
},\rho} - \varepsilon ^{\rho,\widetilde{\omega }}\right) -
\varepsilon ^{0,[\mu\nu]}+\varepsilon ^{[\mu\nu],0} -
\frac{1}{2}e_{\mu\nu\rho\omega}\left(\varepsilon^{\widetilde{0},
[\rho\omega]} -\varepsilon^{[\rho\omega],\widetilde{0}}\right) ,
\]
\[
\frac{1}{2}\overline{\Gamma }_{[\mu} \Gamma _{\nu
]}\equiv\frac{1}{2}\left(\overline{\Gamma }_\mu \Gamma _\nu
-\overline{\Gamma} _\nu \Gamma _\mu\right)= e_{\mu \nu \rho \omega
}\left(\varepsilon ^{\widetilde{\rho},\omega } + \varepsilon
^{\omega,\widetilde{\rho}}\right)
\]
\vspace{-8mm}
\begin{equation}  \label{99}
\end{equation}
\vspace{-8mm}
\[
 + \varepsilon ^{0,[\nu \mu]}+\varepsilon ^{[\nu \mu],0} +
\frac{1}{2}e_{\mu\nu\rho\omega}\left(\varepsilon^{\widetilde{0},
[\rho\omega]} +\varepsilon^{[\rho\omega],\widetilde{0}}\right) .
\]
Using Eqs. (47), we obtain the generators of the Lorentz
transformations in the 16-dimensional space of wave function (20):
\[
J_{\mu \nu }=\frac 14\left( \Gamma _{[\mu} \Gamma _{\nu ]}
+\overline{\Gamma }_{[\mu} \overline{\Gamma }_{\nu ]}\right)
\]
\vspace{-8mm}
\begin{equation}  \label{100}
\end{equation}
\vspace{-8mm}
\[
=\varepsilon ^{[\alpha\mu],[\alpha\nu]}- \varepsilon
^{[\alpha\nu][ \alpha\mu]} +\varepsilon ^{\mu,\nu}-\varepsilon
^{\nu,\mu} + \varepsilon^{\widetilde{\mu},\widetilde{\nu}} -
\varepsilon^{\widetilde{\nu},\widetilde{\mu}} .
\]

\end{document}